# Quantum plasmonic hot-electron injection in lateral $WSe_2/MoSe_2$ heterostructures

Chenwei Tang[1,2], Zhe He[1], Weibing Chen[3], Shuai Jia[3], Jun Lou[3], and Dmitri V. Voronine[4*]

[1] Institute for Quantum Science and Engineering, Texas A&M University, College Station, TX 77843, USA

[2] School of Science, Xi'an Jiaotong University, Xi'an, Shaanxi 710049, China

[3] Department of Materials Science & Nano Engineering, Rice University, Houston, TX 77005, USA

[4] Department of Physics, University of South Florida, Tampa, FL 33620, USA

**Abstract**

Lateral two-dimensional (2D) transitional metal dichalcogenide (TMD) heterostructures have recently attracted a wide attention as promising materials for optoelectronic nanodevices. Due to the nanoscale width of lateral heterojunctions, the study of their optical properties is challenging and requires using subwavelength optical characterization techniques. We investigated the photoresponse of a lateral 2D $WSe_2/MoSe_2$ heterostructure using tip-enhanced photoluminescence (TEPL) with nanoscale spatial resolution and with picoscale tip-sample distance dependence. We demonstrate the observation of quantum plasmonic effects in 2D heterostructures on a non-metallic substrate, and we report the nano-optical measurements of the lateral 2D TMD heterojunction width of ~ 150 nm and the charge tunneling distance of ~ 20 pm. Controlling the plasmonic tip location allows for both nano-optical imaging and plasmon-induced hot electron injection into the heterostructure. By adjusting the tip-sample distance, we demonstrated the controllability of the hot-electron injection via the competition of two quantum plasmonic photoluminescence (PL) enhancement and quenching mechanisms. The directional charge transport in the depletion region leads to the increased hot electron injection, enhancing the $MoSe_2$ PL signal. The properties of the directional hot-electron injection in the quantum plasmonic regime make the lateral 2D $MoSe_2/WSe_2$ heterostructures promising for quantum nanodevices with tunable photoresponse.

[*]Correspondence: voronine@usf.edu

Two-dimensional (2D) transition metal dichalcogenides (TMDs) are promising candidates for optoelectronic devices [1–3], sensors [4,5], and photo-catalysts [6,7]. Assembling two TMD materials vertically or laterally introduced new possibilities [8–11]. Optoelectronic properties of lateral heterostructures are determined by the band structure, doping, and defects of both materials near the boundary [12,13]. Due to the nanoscale size and multi-component optical properties, lateral TMD heterostructures are suitable for single molecule sensing and nano-devices with tunable photoresponse [14–16]. However, in order to fully understand and utilize the unique optoelectronic properties of these 2D materials it is important to characterize and control them with nanoscale spatial resolution.

In tip-enhanced photoluminescence (TEPL), the optical signal is enhanced by focusing light at the apex of the scanning probe (tip) and placing the tip at the close distance d near the sample surface (Fig. 1a). When the tip-sample distance is large enough to ignore tunneling effects, resonant excitation of surface plasmons at the metallic tip generates large electromagnetic field enhancement, and therefore provides photoluminescence (PL) enhancement and increase in spatial resolution [17], [18–21]. Apart from this electromagnetic mechanism, hot carriers can also be injected into the sample and contribute to the improved sensitivity and spatial resolution via the charge transfer mechanism [22,23]. When the distance between the tip and the substrate decreases to the "quantum regime", i.e. sub-nanometer range, electron tunneling becomes significant, leading to the attenuation of the local electromagnetic field and thus the quenching of the emission [24–33]. This tunneling-induced quenching of the PL signal is the optical signature of quantum plasmonics. High spatial resolution of tip-enhanced imaging has recently been demonstrated [34–41]. The spatial resolution depends on the near field enhancement which may be optimized by varying the tip-sample distance. Plasmon-induced hot electron injection may also contribute to the PL enhancement via the carrier recombination mechanism.

Hot electrons are carriers with high kinetic energies that can be generated using bias [42], photocurrent [43], and plasmonic nanostructures including metallic tips [44–46]. Injecting hot electrons into TMDs may facilitate photocatalytic reactions [6,47] and photoelectron emission [48]. Plasmonic tips are especially suitable for hot electron generation and injection in TMDs due to the strong local electromagnetic fields [49] which may be described using the metal-semiconductor coupling model [50–53]. Moreover, directional hot electron transfer in TMD heterostructures may take place in the depletion region [50–52] and may be enhanced by adjusting the tip-sample distance in the sub-nanometer quantum regime. Quantum plasmonic effects were previously observed in plasmonic metallic nanostructures with sub-nanometer gaps, due to the tunneling-induced quenching of the local electric fields [24–33]. Since the tunneling is sensitive to the gap size, it is possible to manipulate it by varying the gap size at the picometer scale from the classical regime where there is no tunneling to the quantum regime there the tunneling makes a significant contribution to the field quenching. Recently, we developed a picometer-scale tip-sample distance dependence approach which can be used to perform such precise measurements [30]. This approach is well suited for studying quantum plasmonic effects in TMD heterostructures.

In this work, we investigate tunneling-assisted hot electron injection (HEI) in the system of coupled Au-coated plasmonic Ag tip and $MoSe_2$-$WSe_2$ heterostructure on a Si/$SiO_2$ substrate at room temperature. We observed quenching and enhancement of the PL from the 2D heterojunction due to the attenuation of localized electromagnetic field and hot electron injection, respectively. Using the near-field TEPL imaging, we performed nanoscale optical characterization of the heterojunction and achieved control of the junction PL by varying the nanoscale lateral tip position and picoscale tip-sample distance. For the tip-sample distance d > 0.36 nm, the classical plasmon-induced hot electron injection is limited by the air gap barrier. For gap sizes comparable to or smaller than the van der Waals (vdW) contact distance, the electron tunneling facilitates thermionic injection in the quantum regime [53]. This provides an alternative mechanism for manipulating the optoelectronic properties of 2D materials which may be used for improving the characterization and design of TMD based devices.

The monolayer lateral type II $MoSe_2$-$WSe_2$ heterostructure was grown on the Si/$SiO_2$ substrate via chemical vapor deposition [54,55]. $MoO_3$ and $WO_3$ acting as precursors were placed into the center of the furnace. Se powder at upstream was introduced into the furnace center by the hydrogen gas and it reacted with $MoO_3$ and $WO_3$ precursors to grow $MoSe_2$-$WSe_2$ heterojunctions at 750 ℃. Atomic force microscopy (AFM) revealed uniform thickness of the selected triangle area of < 2 nm (Fig. 1b). We compared the photoluminescence (PL) intensity of the heterostructure with the Raman intensity of the Si/$SiO_2$ substrate to determine the number of layers. The near unity ratio and the narrow full width half maximum (FWHM) of the PL signal strongly indicate that the sample is a monolayer. The chemical composition of the $MoSe_2$ and $WSe_2$ parts of the heterostructure was also confirmed using the PL and Raman spectra shown in Figs. 1c and 1d, respectively. Note that the PL peak position and FWHM of monolayer $WSe_2$ and $MoSe_2$ vary in previous reports [56–59]. This may be due to the strain or defects.

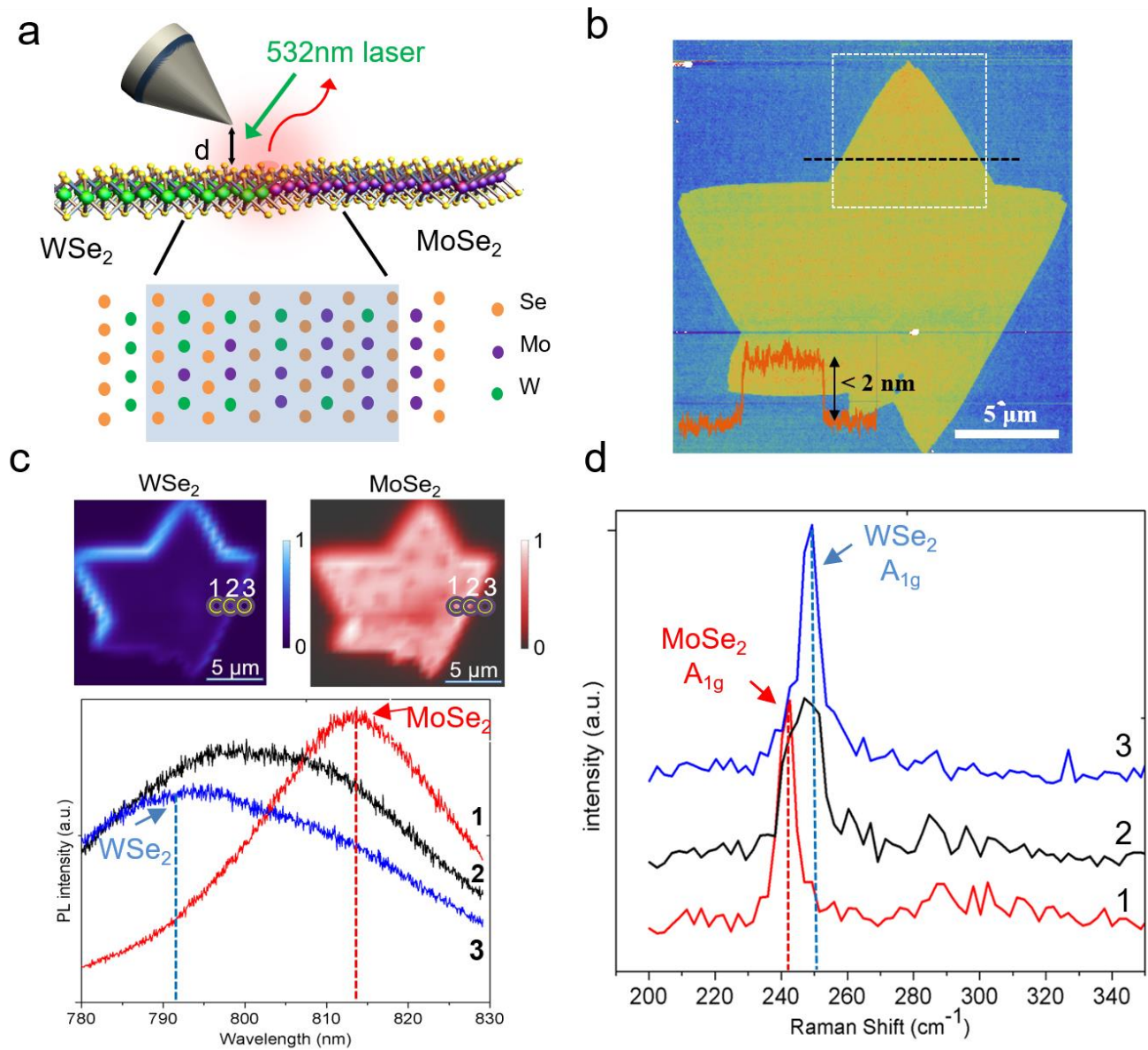

Figure 1. Lateral 2D MoSe$_2$-WSe$_2$ heterostructure. (a) Sketch of the tip-enhanced photoluminescence (TEPL) measurement setup. 532 nm linearly polarized laser (green arrow) is focused onto a Au-coated plasmonic Ag nanotip operated in the contact mode with the controllable tip-sample distance d. The back-scattered TEPL signal (red arrow) is collected as a function of d in the classical (d > 0.36 nm) and quantum plasmonic (d < 0.36 nm) regimes. (b) Atomic force microscopy (AFM) image of the MoSe$_2$-WSe$_2$ heterostructure. The bottom profile curve shows a uniform sample thickness < 2 nm along the black dashed line. (c) Normalized far-field PL images of the WSe$_2$ (blue) and MoSe$_2$ (red) parts of the heterostructure. Highlighted spots 1, 2, and 3 correspond to the MoSe$_2$, junction and WSe$_2$ parts of the heterostructure, respectively. (d) The corresponding far-field Raman spectra show the peaks of WSe$_2$ at 250 cm$^{-1}$ [60], MoSe$_2$ at 242 cm$^{-1}$ [61] and both peaks at the heterojunction.

TEPL was carried out using the state-of-the-art commercial system (OmegaScope-R coupled with LabRAM Evolution microscope, Horiba Scientific). Silicon tips with apex radius ~ 10 nm were used for AFM. Au-coated Ag tips with apex radius ~ 20 nm were used for TEPL measurements. Therefore, the spatial resolution of TEPL was ~ 40 nm as determined by the tip diameter. The 532 nm linearly polarized laser radiation was focused onto the tip apex at an incident angle of 53 degrees and the resulting PL signals were collected using the same objective (100×, NA 0.7, f = 200). The sample was scanned while recording at each point both the near-field and the far-field signals with the controllable tip-sample distance d ~ 0.36 nm and ~ 20 nm, respectively. For the results shown in Figs. 2a and 2b, the laser power was 2.5 mW and the sample scanning step size was 40 nm, with 0.2 s acquisition time. The results shown in Figs. 2d and 2e were obtained with the same laser power and acquisition time while the scanning step size was 1 nm.

We performed TEPL imaging of the part of the $MoSe_2$-$WSe_2$ heterostructure marked by the dashed rectangular area in Fig. 1b. The double Gaussian fitting of the heterojunction PL is shown in Fig. 2c, where the $WSe_2$ (centered at 783 nm, 1.58 eV) and $MoSe_2$ (centered at 806 nm, 1.54 eV) components are shown by blue and red curves, respectively. The integrated values within FWHM of these Gaussian functions represent the total PL intensities of both components. Figs. 2a and 2b show the integrated PL intensity distributions of the $WSe_2$ and $MoSe_2$ for the near-field TEPL and far-field PL, respectively. The near-field image in Fig. 2a shows a sharper heterojunction boundary than the far-field image in Fig. 2b. To estimate the junction width, we scanned the tip along the white dashed line across the junction with 1 nm step size (Figs. 2d and 2e), which showed ~150 nm and >400 nm width in the near-field and far-field profiles, respectively. Therefore, compared with the confocal PL microscopy, TEPL with subwavelength spatial resolution is more suitable for probing the spectral properties of the heterojunction.

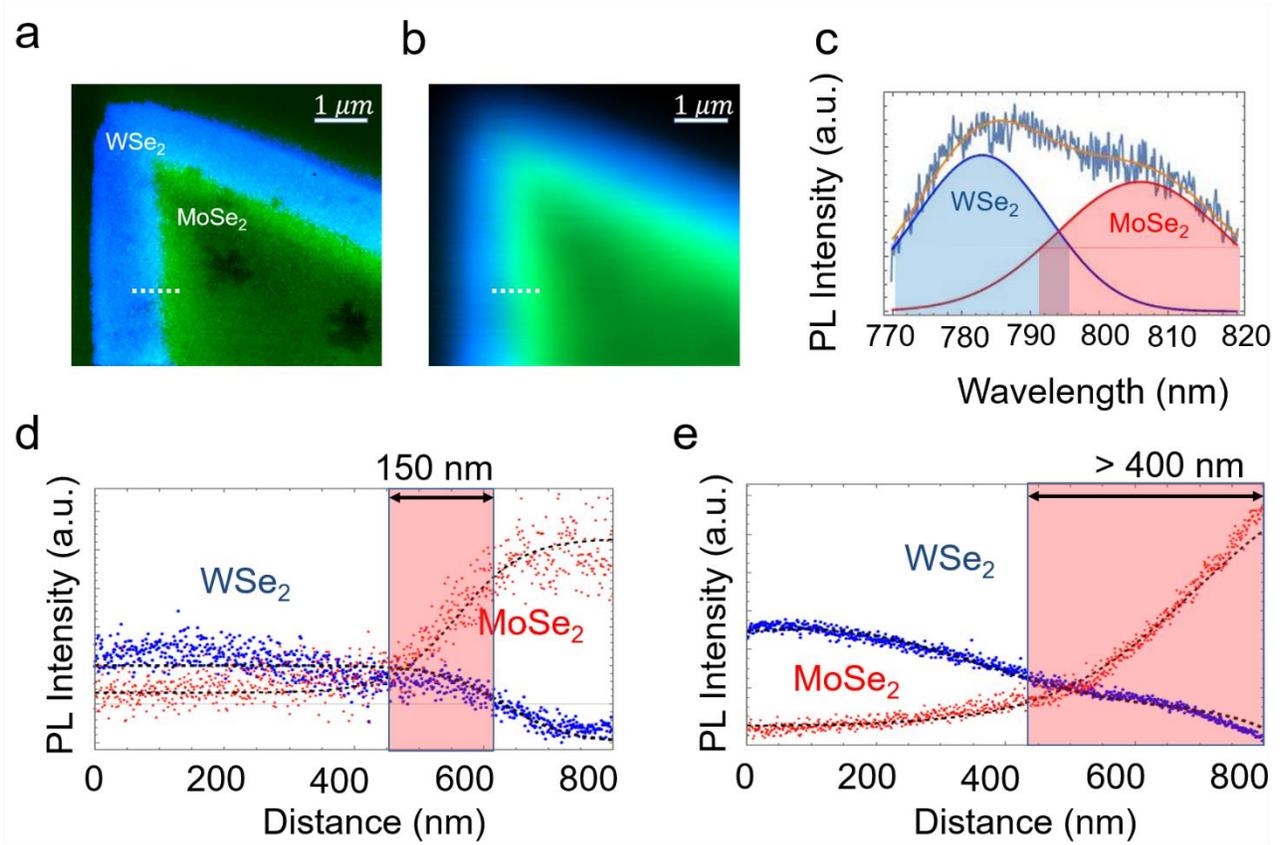

Figure 2. Tip-enhanced photoluminescence (TEPL) imaging of the MoSe$_2$-WSe$_2$ heterostructure. (a) Near-field PL image with the tip-sample distance d ~ 0.36 nm and (b) far-field PL image with d ~ 20 nm. The green and blue areas correspond to the integrated MoSe$_2$ (806 nm) and WSe$_2$ (783 nm) PL signals, respectively. The PL intensity of each component is obtained by integrating the area which corresponds to the FWHM of each component's Gaussian fit (c). Spatial dependence of the near-field (d) and far-field (e) PL intensity of both components along a white dashed line crossing the heterojunction marked in (a) and (b), respectively. The heterojunction width is highlighted in (d) and (e) by the shaded red areas.

To understand the effects of the tip-sample interaction on the heterostructure PL, we varied the tip-sample distance from ~ 40 nm to the Au-S vdW contact of ~ 0.36 nm and further down to ~ 0.2 nm which was estimated according to the contact force via the Lennard-Jones potential [18,22,30,31,35]. The picometer-scale tip-sample distance dependence calibration procedure was used as previously described [30]. Two main factors, namely, the local electromagnetic field and the hot electron injection, contribute to the tip-sample distance dependence of the PL signal. As the tip-sample distance decreases from 40 nm to 20 nm, the enhanced electromagnetic field increases while the hot electron injection can be neglected (Figs. 3 and 4). For pure WSe$_2$, the near-field enhancement saturates at d ≈ 20 nm, at which point, the hot electron injection rate increases, leading to a competition between the PL enhancement induced by hot electrons and quenching due to the attenuation of the tip electric field [53]. For the CVD-grown pure $WSe_2$, the concentration of the holes is larger than the intrinsic electrons, and, therefore, its PL shows gradual enhancement in the classical regime of 0.36 nm < d < 40 nm (Figs. 3a and 3c). Once the tip and $WSe_2$ are within the vdW contact separation (d = 0.36 nm), thermionic injection may occur with the increasing electron density in $WSe_2$ [53]. Consequently, an abrupt increase of the PL intensity in pure $WSe_2$ can be seen in Figs. 3a and 3c. On the other hand, for the CVD-grown $MoSe_2$, as the electron-hole recombination is limited by the lack of the intrinsic holes, tunneling is suppressed and, therefore, the PL of pure $MoSe_2$ shows no significant enhancement at 0.36 nm (Figs. 3b and 3c). Compared with the results from the heterojunction shown in Fig. 4a, no significant quenching of pure $WSe_2$ PL and no enhancement of pure $MoSe_2$ PL was observed.

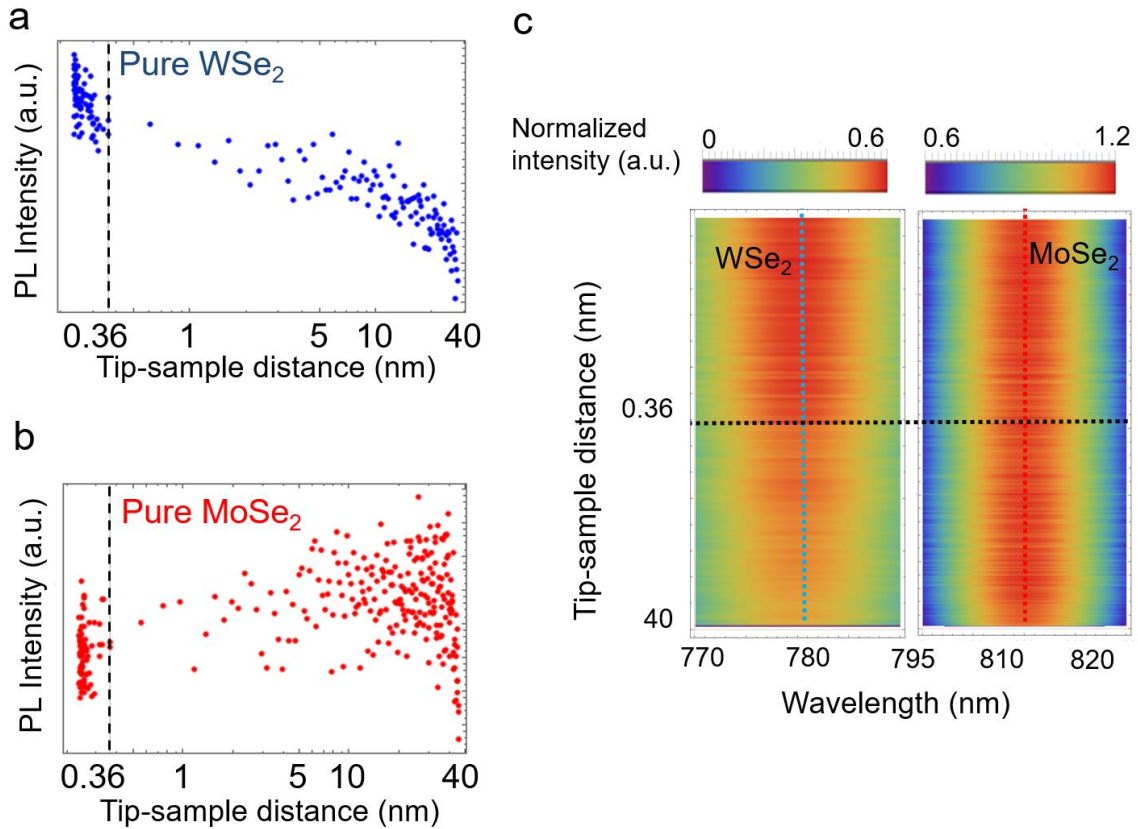

Figure 3. Distance-dependence of the photoluminescence (PL) intensity of the pure monolayer 2D materials: (a) $WSe_2$ and (b) $MoSe_2$. (a) PL of $WSe_2$ increases when the tip-sample distance decreases from 40 nm to 0.26 nm due to the classical plasmonic enhancement. (b) PL of pure $MoSe_2$ does not show significant enhancement both in the classical (d > 0.36 nm) and quantum (d < 0.36 nm) regimes under the similar experimental conditions as $WSe_2$

due to the difference in the PL and tunneling efficiencies. (c) Distance-dependence of the PL spectra of the pure monolayer $WSe_2$ and $MoSe_2$.

At the $MoSe_2/WSe_2$ heterojunction, directional hot electron injection may take place due to the carrier-deficient depletion region formed at the junction [50–52]. Fig. 4 shows a schematic energy diagram of the hot electrons transferred to $MoSe_2$ due to the chemical potential gradient at the heterojunction [8,9]. As the tip diameter is comparable to or narrower than the size of the $WSe_2$-$MoSe_2$ depletion region, the energy needed for the diffused hot electrons transferred to the $MoSe_2$ side to tunnel back to the tip is higher because of the air barrier. The Au-semiconductor depletion region [50] also performs as a barrier to reduce the backward tunneling of the dissipated hot electrons in the $MoSe_2$. Therefore, as the tip-sample distance decreases from 20 nm to 0.36 nm, the injected hot electrons accumulate in $MoSe_2$, resulting in the PL enhancement in $MoSe_2$ while quenching the PL in $WSe_2$ at the heterojunction (Figs. 4c and 4d) [48], [62]. It is the depletion region that allows for the $MoSe_2$ side of the heterojunction accumulating more plasmon-induced hot electrons than in the pure materials. This directionality contributes to the unequal enhancement of the TEPL signals from the $MoSe_2$ and $WSe_2$ sides of the heterojunction.

Once the tip-sample distance reaches the vdW contact (d = 0.36 nm), the sub-nanometer gap between the tip and the sample leads to the electron tunneling. Previous TEPL measurements in sub-nanometer gap metal-metal contacts showed that the PL is quenched due to tunneling [31]. Despite the absence of the metal-metal contacts here, the PL intensity of the $WSe_2$ component at the heterojunction also shows quenching (Figs. 4a, 4c, and 4e). The repeated TEPL measurements on the heterostructure confirm the observed effects. On the other hand, the $MoSe_2$ component shows abrupt enhancement when d < 0.36 nm (Figs. 4a, 4d, and 4f). These phenomena can be explained by the decrease of the air barrier between the tip and the sample, leading to the increased number of hot electrons injected into $MoSe_2$ due to tunneling and the corresponding decrease of the surface charge density and near-field intensity at the tip [31,33]. The HEI enhances the PL signal of the $MoSe_2$ part of the junction due to the increase of the recombination rate which is larger than the PL decrease due to the near-field quenching mechanism. On the other hand, the hot electron accumulation in the $WSe_2$ part of the junction is suppressed due to the charge transfer across the depletion region leading to the overall quenching of the $WSe_2$ PL. This delicate interplay of the two PL enhancement mechanisms may be controlled by the lateral tip position and tip-sample distance dependence. Due to the nanoscale size, the plasmonic tip can be used to generate hot electrons with high precision in the depletion region formed by the heterostructure which may be used for designing controllable nano-devices.

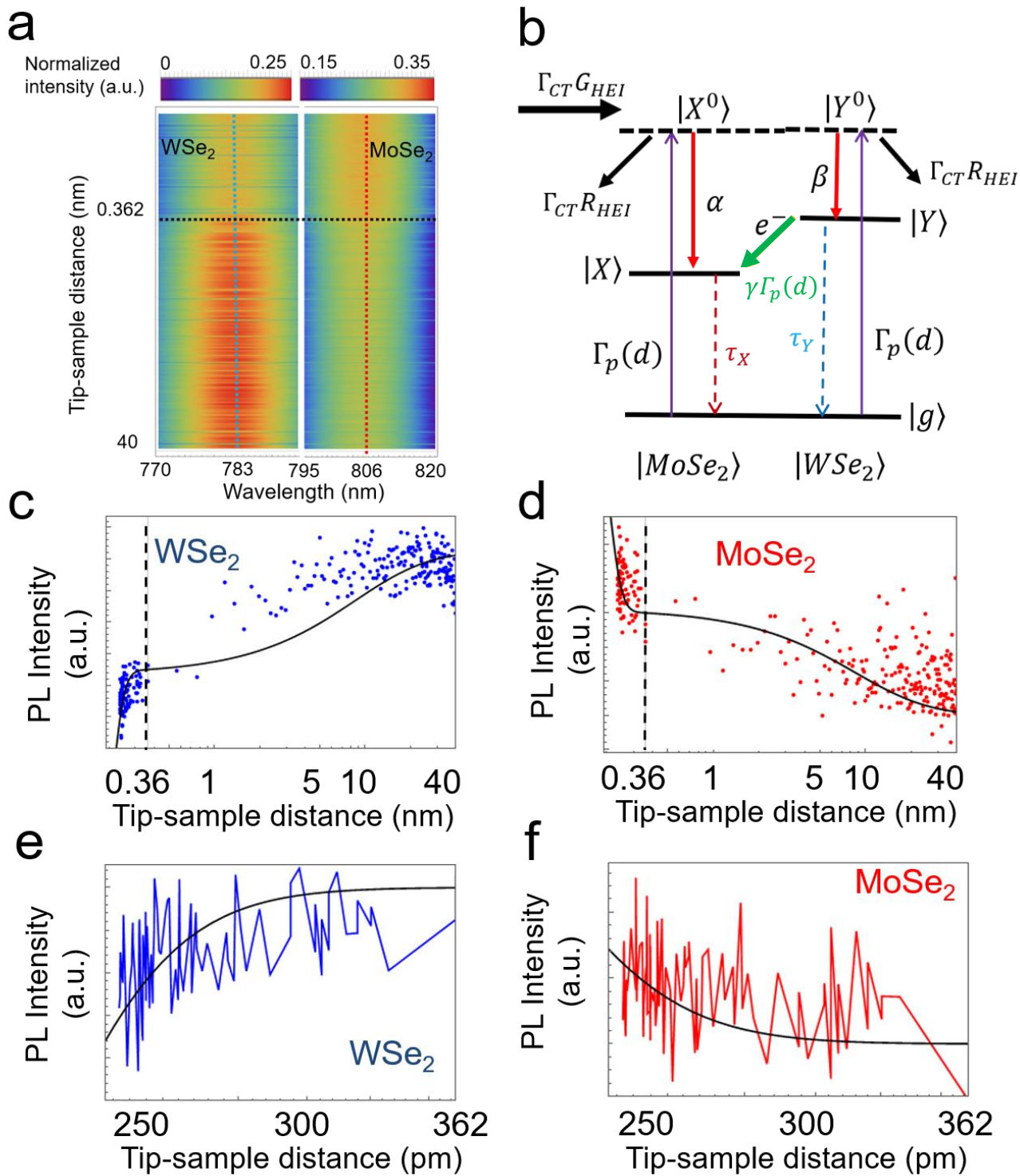

Figure 4. Directional plasmonic hot electron injection in the WSe$_2$-MoSe$_2$ heterostructure revealed by TEPL distance dependence. (a) TEPL tip-sample distance dependence from 40 nm to 0.2 nm, showing abrupt changes in the PL spectra of WSe$_2$ and MoSe$_2$ components when the tip-sample distance d ≤ 0.36 nm, indicating the quantum-to-classical transition in the photo-response of the heterostructure coupled to the plasmonic tip. (b) Energy diagram of the lateral MoSe$_2$-WSe$_2$ heterojunction shows the directional hot electron injection due to the potential gradient at the junction. The control mechanisms are shown: hot electron injection (thick black arrows) and plasmon-induced charge transfer (green arrow), and TEPL (purple) leading to the controllable quenching or enhancement of the PL signals (dashed arrows). (c, d) Tip-sample distance dependence of the PL intensities of the WSe$_2$ and MoSe$_2$ components at the junction. The dashed line at d = 0.36 nm corresponds to the van der Waals (vdW) contact distance between the tip and the sample with the transition from the classical to the quantum regime. (e, f) Zoomed-in plots of the tip-sample PL distance dependence at the junction in the quantum regime. Solid black lines in (c) – (f) are the fittings obtained using the theoretical model described below.

Next, we present a theoretical model used to fit the data which accounts for the competition between the two quantum plasmonic effects, explaining the tunneling-induced HEI and near field PL quenching mechanisms. Fig. 4b shows the simplified diagram of the energy states and transitions at the WSe$_2$-MoSe$_2$ junction, including the effects of the HEI, photo-induced charge transfer, and TEPL due to the tip-sample interaction. The simulation results are shown in Figs. 4c-4f (black curves). The initial state populations $N_{X0}$, $N_{Y0}$, and $N_g$ of the excited states $|X^0\rangle$, $|Y^0\rangle$, and the ground state $|g\rangle$, respectively, and the exciton populations $N_X$, and $N_Y$ of the MoSe$_2$ state $|X\rangle$ and the WSe$_2$ state $|Y\rangle$, respectively, can be described by the rate equations [63]:

$$\frac{dN_{X0}}{dt} = (G_{HEI} - R_{HEI}N_{X0})\Gamma_{CT}(d) - \alpha N_{X0} + \Gamma_p(d)(N_g - N_{X0}), \quad (1)$$

$$\frac{dN_X}{dt} = \alpha N_{X0} + \gamma \Gamma_p(d)N_Y - \frac{N_X}{\tau_X}, \quad (2)$$

$$\frac{dN_{Y0}}{dt} = -\beta N_{Y0} - R_{HEI}N_{Y0}\Gamma_{CT}(d) + \Gamma_p(d)(N_g - N_{Y0}), \quad (3)$$

$$\frac{dN_Y}{dt} = \beta N_{Y0} - \gamma \Gamma_p(d)N_Y - \frac{N_Y}{\tau_Y}, \quad (4)$$

$$\frac{dN_g}{dt} = -\Gamma_p(d)(N_g - N_{X0}) - \Gamma_p(d)(N_g - N_{Y0}) + \frac{N_X}{\tau_X} + \frac{N_Y}{\tau_Y}, \quad (5)$$

where $\Gamma_{CT}G_{HEI}$ is the hot electron injection (HEI) rate, and $\Gamma_{CT}R_{HEI}$ is the hot electron decay rate from states $|X^0\rangle$ or $|Y^0\rangle$ [63]. We assume $G_{HEI} = R_{HEI} = 1$. The tunneling $\Gamma_{CT}(d)$ is given by:

$$\Gamma_{CT}(d) = \begin{cases} Ae^{-\frac{d-c}{d_{CT}}}, & \text{for } d < 0.36 \, nm, \\ 0, & \text{for } d > 0.36 \, nm, \end{cases} \quad (6)$$

where $d_{CT}$ is the average tunneling distance in the quantum regime. For simplicity, we neglected the tunneling for the tip-sample distance larger than the vdW contact distance of 0.36 nm. $A$ is the normalization parameter. The near-field pumping rate $\Gamma_p(d)$ describes the local optical excitation by the near field of the tip as [18]

$$\Gamma_p(d) = \begin{cases} 1 - e^{-\frac{d-c}{d_p}}, & \text{for } c < d < 0.36 \, nm, \\ B(R + d - c)^{-4}, & \text{for } d > 0.36 \, nm, \end{cases} \quad (7)$$

where $d_p$ is the average quantum coupling distance, which leads to the quenching of the optical excitation by the tunneling in the quantum regime. R = 25 nm is the tip radius, B is a fitting parameter to smoothen the piecewise function, and c is the conductive contact distance which corresponds to the Ohmic tip-sample contact. When the tip-sample distance d < c, the near field of the tip is completely quenched and the pumping rate $\Gamma_p(d) = 0$. Since TEPL depends on $\Gamma_p(d)$, we only consider the region of d > c. The exciton generation rates of MoSe$_2$ and WSe$_2$ are given by $\alpha = \beta = 1 \, ps^{-1}$, respectively [64]. $\gamma\Gamma_p(d)$ is the photo-induced charge transfer rate across the junction which was assumed to be proportional to the near-field pumping rate, where we assume $\gamma = 1$. The near-field excitation may facilitate the charge transfer within heterostructures [65]. The exciton relaxation times of MoSe$_2$ and WSe$_2$ were taken as $\tau_X = \tau_Y = 2 \, ps$ [66].

The model was used to fit the tip-sample distance dependence results shown in Fig. 4. The main fitting parameters were $d_{CT} = d_p = 0.02 \, nm$, and c = 0.17 nm. The latter shows the measured conductive contact distance approximately equal to the half of the vdW contact distance (0.36 nm) between the Ag and S atoms. If the tip-sample distance d is decreased to ~ 0.17 nm, the near-field TEPL signal is expected to be completely quenched. In our experiments, the shortest tip-sample distance was ~ 0.20 nm and did not reach 0.17 nm due to the possible tip damage. The sub-nm values of the mean distance parameters $d_{CT}$ and $d_p$ reflect the tunneling nature of the quantum plasmonic effects. Our model fits the experimental results well in both the classical (d > 0.36 nm) and quantum regimes (d < 0.36 nm) as shown in Figs. 4c and 4d. In the classical regime, since the near-field pumping rate $\Gamma_p(d)$ is larger when the distance d approaches 0.36 nm, the photo-induced charge transfer $\gamma\Gamma_p(d)$ leads to the decrease of the WSe$_2$ PL signal and the increase of the MoSe$_2$ PL signal. On the other hand, in the quantum regime, the hot electron injection and PL quenching effects are dominant. The main goal of our work was to investigate the effects of HEI on TEPL in a TMD heterostructure. Our simplified model is sufficient to describe the two main tunneling-induced effects, namely, HEI PL enhancement and TEPL quenching. The model could be further improved by including the electron and hole densities, atomic orbital contributions of the electron and hole states, and band offsets, which is, however, beyond the scope of the current paper.

## Acknowledgments

C.T. acknowledges the support from the Education Program for Talented Students of Xi'an Jiaotong University. Z.H is supported by the Herman F. Heep and Minnie Belle Heep Texas A&M University Endowed Fund held/administered by the Texas A&M Foundation. We thank Prof. Marlan Scully for the use of laser facilities. D.V.V. acknowledges the support of the National Science Foundation under Grant CHE-1609608. W.C., S.J, and J.L. acknowledge the support of the Welch Foundation (Grant C-1716).